\catcode`\@=11

\newif\if@fewtab\@fewtabtrue

{\count255=\time\divide\count255 by 60 \xdef\hourmin{\number\count255}
\multiply\count255 by-60\advance\count255 by\time
\xdef\hourmin{\hourmin:\ifnum\count255<10 0\fi\the\count255}}
\def\ps@draft{\let\@mkboth\@gobbletwo
    \def\@oddhead{}
    \def\@oddfoot
       {\hbox to 7 cm{$\scriptstyle Draft\ version:\ \draftdate$
       \hfil}\hskip -7cm\hfil\rm\thepage \hfil}
    \def\@evenhead{}\let\@evenfoot\@oddfoot}

\def\ceqno{\global\@fewtabfalse
    \ifcase\@eqcnt \def\@tempa{& & &}\or \def\@tempa{& &}
      \or \def\@tempa{&}
      \or\def\@tempa{}\fi\@tempa
{\rm(\theequation)}}

\def\aeqno#1{\global\@fewtabfalse
    \ifcase\@eqcnt \def\@tempa{& & &}\or \def\@tempa{& &}
      \or \def\@tempa{&}
      \or\def\@tempa{}\fi\@tempa
{\rm(\theequation,#1)}}

\def\label#1{\ifnum\draftcontrol=1
 \global\def\draftnote{$\scriptstyle #1$}\fi
 \@bsphack\if@filesw {\let\thepage\relax
   \def\protect{\noexpand\noexpand\noexpand}%
\xdef\@gtempa{\write\@auxout{\string
      \newlabel{#1}{{\@currentlabel}{\thepage}}}}}\@gtempa
   \if@nobreak \ifvmode\nobreak\fi\fi\fi
  \@esphack}

\def\alabel#1#2{\label{#1}\global\@fewtabfalse
    \ifcase\@eqcnt \def\@tempa{& & &}\or \def\@tempa{& &}
      \or \def\@tempa{&}
      \or\def\@tempa{}\fi\@tempa
{\hbox to 3cm{\phantom{\rm(\theequation,#2)}
\draftnote \hfil}\hskip -3cm {\rm(\theequation,#2)}}}

\def\clabel#1{\label{#1}\global\@fewtabfalse
    \ifcase\@eqcnt \def\@tempa{& & &}\or \def\@tempa{& &}
      \or \def\@tempa{&}
      \or\def\@tempa{}\fi\@tempa
{\hbox to 3cm{\phantom{\rm(\theequation)}
\draftnote \hfil}\hskip -3cm{\rm(\theequation)}}}

\def\eqnarray{\def\draftnote{{}}\global\@fewtabtrue
\stepcounter{equation}\let\@currentlabel=\theequation
\global\@eqnswtrue
\global\@eqcnt\z@\tabskip\@centering\let\\=\@eqncr
$$\halign to \displaywidth\bgroup\@eqnsel\hskip\@centering\@eqcnt\z@
  $\displaystyle\tabskip\z@{##}$&\global\@eqcnt\@ne
  \hskip 1\arraycolsep \hfil${##}$\hfil
  &\global\@eqcnt\tw@ \hskip 1\arraycolsep $\displaystyle\tabskip\z@{##}$
\hfil  \tabskip\@centering&\global\@eqcnt\thr@@\llap{##}\tabskip\z@
\cr}

\def\endeqnarray{\@@eqncr\egroup
      \global\advance\c@equation\m@ne$$\global\@ignoretrue}

\def\@eqnnum{\hbox to 3cm{\phantom{\rm(\theequation)} \draftnote
                         \hfil}\hskip -3cm {\rm(\theequation)}}

\def\@@eqncr{\let\@tempa\relax
    \ifcase\@eqcnt \def\@tempa{& & &}\or \def\@tempa{& &}
      \or \def\@tempa{&}
      \or\def\@tempa{}
\fi\@tempa
\if@eqnsw
\if@fewtab\@eqnnum\fi
\stepcounter{equation}\fi\global
\@eqnswtrue\global\@eqcnt\z@\global\@fewtabtrue\cr}

\def\draftcite#1{\ifnum\draftcontrol=1#1\else{}\fi}

\def\@lbibitem[#1]#2{\item{}\hskip -3cm \hbox to 2cm
{\hfil$\scriptstyle\draftcite{#2}$}\hskip 1cm[\@biblabel{#1}]\if@filesw
     {\def\protect##1{\string ##1\space}\immediate
      \write\@auxout{\string\bibcite{#2}{#1}}}\fi\ignorespaces}

\def\@bibitem#1{\item\hskip -3cm \hbox to 2cm
{\hfil $\scriptstyle\draftcite{#1}$}\hskip 1cm
\if@filesw \immediate\write\@auxout
       {\string\bibcite{#1}{\the\value{\@listctr}}}\fi\ignorespaces}

\font\tendl=msym10  scaled \magstep1
\font\sevendl=msym7 scaled \magstep1
\font\fivedl=msym5 scaled \magstep1
\font\tengl=eufm10  scaled \magstep1
\font\sevengl=eufm7 scaled \magstep1
\font\fivegl=eufm5 scaled \magstep1

\newfam\dlfam \def\dl{\fam\dlfam\tendl} 
\textfont\dlfam=\tendl \scriptfont\dlfam=\sevendl
\scriptscriptfont\dlfam=\fivedl
\newfam\glfam  
\textfont\glfam=\tengl \scriptfont\glfam=\sevengl
\scriptscriptfont\glfam=\fivegl

\def\draftdate{\number\month/\number\day/\number\year\ \ \ \hourmin }

\global\def\draftcontrol{0}
\catcode`\@=12

\def\hat{\widehat}
\documentstyle[12pt]{article}

\def\theequation{{\arabic{equation}}}

\newcommand{\be}{\begin{eqnarray}}
\newcommand{\en}{\end{eqnarray}\vs 0.5 cm}

\newcommand{\no}{\noindent}
\newcommand{\vs}{\vskip}
\newcommand{\hs}{\hspace}

\newcommand{\p}{\partial}
\newcommand{\un}{\underline}
\newcommand{\NR}{{{\dl R}}}
\newcommand{\NC}{{{\dl C}}}
\newcommand{\NT}{{{\dl T}}}
\newcommand{\NZ}{{{\dl Z}}}
\newcommand{\NN}{{{\dl N}}}
\newcommand{\NH}{{{\dl H}}}
\newcommand{\qq}{\begin{eqnarray}}

\newcommand{\ee}{{\rm e}}

\newcommand{\qqq}{\end{eqnarray}}

\newcommand{\CA}{{\cal A}}

\newcommand{\CG}{{\cal G}}

\newcommand{\s}{\hspace{0.05cm}}
\def\pref#1{(\ref{#1})}
\begin{document}
\centerline{\bf CHERN-SIMONS THEORY ON THE TORUS\footnote{talk
given by the first author at the
XIXth International Colloquium on Group
Theoretical Methods in Physics, Salamanca (Spain),
June 29-July 4, 1992}}
\centerline{F.Falceto\footnote{Address after October 1$^{\rm st}$, 1992: Depto.
F\'{\i}sica Te\'orica, U. Zaragoza, 50009 Zaragoza, Spain}, K.Gaw\c{e}dzki}
\centerline{IHES, Bures-sur-Yvette, 91440,
France}
\hskip .5cm


Recently a considerable effort has been invested in  understanding
Chern-Simons theories from the canonical or covariant points of view.
The covariant quantization was used to obtain
topological invariants of three
dimensional manifolds \cite{Witten}  whereas the canonical point
of view [1-5] allowed to relate the states of three dimensional
Chern-Simons theory to conformal blocks of two dimensional
Wess-Zumino-Witten models.

Here we shall discuss the second approach.
More precisely we shall study the space of quantum states
of the Chern-Simons theory for the group $SU(2)$ on $\NT^2\times\NR$
and in the presence of Wilson lines $\{z_n\}\times\NR,\ n=1,\cdots,N,$
corresponding to representations $\rho_{j_n}$ of spin $j_n$
(acting in spaces $V_{j_n}$).

The gauge freedom is partially fixed by setting the temporal ($\NR$-direction)
component of the connection to zero. The remaining gauge
invariance is imposed on the states
(very much like the Gauss law in QED) in terms of a
quantum flatness condition.
The presence of the Wilson lines contributes source terms to the flattness
condition.

In the holomorphic quantization {\it \`a la Bargmann}
we pick up a complex structure
on the torus $\NT^2$ by fixing the modular parameter
$\tau=\tau_1+i\tau_2\in\NH$. This induces a complex structure
in the space of two dimensional (smooth) connections.
Integration of the flattness condition leads then to the following picture
of the quantum states (see \cite{G}, \cite{GK}, \cite{fg},):
they are holomorphic functionals
$$\Psi:\CA^{01}\equiv\Omega^{01}(T^2)
\otimes{\rm sl}(2,\NC)\rightarrow\bigotimes_nV_{j_n}$$
satisfying
\qq
\Psi({}^h{\hspace{-1mm}}A^{01})=
\ee^{kS_{W\hspace{-0.05 cm}ZW}(h^{-1},A^{01})}
\prod_n h(z_n)_{(n)}\s\Psi(A^{01}),
\label{blocks}
\qqq
where $S_{W\hspace{-0.05 cm}ZW}$ is the Wess-Zumino-Witten
action coupled to the $(0,1)$ part of the connection whose
gauge transformed version ${}^h{\hspace{-1mm}}A^{01}$
is given by
\qq
{}^h{\hspace{-1mm}}A^{01}=hA^{01}h^{-1}+
h\bar\p h^{-1}\label{gauge}
\qqq
with
$h$
in the gauge group $\CG^\NC\equiv C^\infty(\NT^2,{\rm SL}(2,\NC))$.
$h(z)_{(n)}\equiv I\otimes\cdots\otimes\rho_{j_n}(h(z))
\otimes\cdots\otimes I$.

{}From \pref{blocks} it follows that $\Psi$ is determined from
its value at a point of each gauge orbit.
However the space $\CA^{01}/\CG^\NC$ is not a manifold
and to study the smoothness of $\Psi$ subject to
\pref{blocks} we need a detailed knowledge of positions
and sizes (codimensions) of the $\CG^\NC$ orbits.

First we shall identify $\CA^{01}$ with the space
of structures of holomorphic $SL(2,\NC)$ bundles
on the trivial bundle $\NT^2\times\NC^2$: holomorphic sections
of the bundle corresponding to $A^{01}$ are given by maps $g$ with
$(\bar\partial+A^{01})g=0$. This space has been
studied in \cite{AB} and from there we infer that
in our case almost every bundle, except for some submanifolds
of complex codimension at least $2$, is semistable
and all these strata are formed by the union of gauge
orbits. From the properties of the stratification and the
Hartogs theorem, it follows that every
holomorphic functional on the space of semistable bundles can be uniquely
extended to the whole space.

Every semistable connection can be
gauge transformed into one of
the following \cite{Gunn}\cite{fg}
\qq\nonumber
A^{01}_u&=&-u\sigma_3 d\bar z/(2\tau_2)&$u\in\NC$\hbox to 3.75cm{}\cr
\hat A_\alpha^{01}&=&{}^{h_\alpha}(-\sigma_+d\bar z/(2\tau_2))
&$\alpha=0,{{1\over2}},{{\tau\over2}}, {{\tau+1\over2}},$\hbox to 2cm{}\cr
\qqq
\vskip-\baselineskip\no
with $h_\alpha=\exp[(\alpha\bar z-\bar\alpha z)\sigma_3/(2\tau_2)]$. Note that
for $\alpha={{1/2}},{{\tau/2}},
{{(\tau+1)/2}}$, $h_\alpha\not\in\CG^\NC$, however its gauge action
on $\CA^{01}$ is still well defined by \pref{gauge}.

The set $\CA^{01}_0$, formed by the
union of $\CG^\NC$ orbits of
$A^{01}_u$ for $u\in\NC\setminus(\NZ+\tau\NZ)/2$,
is open dense in $\CA^{01}$, while
orbits $\CA^{01}_{(\alpha,0)}\equiv{}^{\CG^\NC} \hat A_\alpha^{01}$ and
$\CA^{01}_{(\alpha,1)}\equiv{}^{\CG^\NC} A_\alpha^{01}$ are for
$\alpha=0,1/2,\tau/2,(\tau+1)/2$ complex submanifolds
of codimension $1$ and $3$ respectively (see \cite{fg} for
details and proofs).

{}From every holomorphic functional $\Psi$ on
$\CA^{01}$ we define a holomorphic function
$\gamma:\NC\rightarrow\bigotimes_nV_{j_n}$
by
\qq
\Psi(A_u^{01})\equiv e^{\pi k u^2/\tau_2}
\prod_n (e^{\sigma_3(\bar z_n-z_n)u/(2\tau_2)})_{(n)}
\ \gamma(u)\hspace{.06cm}.\label{blocks1}
\qqq
If  $\Psi$ satisfies \pref{blocks}
then
\qq
\gamma(u+1)\hspace{0.04 cm}&=&\gamma(u),\aeqno a\cr
\ &\cr
\gamma(u+\tau)&=&e^{-2\pi i k  (\tau +2 u)}\hspace{.06 cm}
\prod_{l} (e^{i z_n\sigma_3})_{(n)}\ \gamma(u),\alabel{cond}b\cr
\ &\cr
0\ \ \ \ \ \ &=&\sum_{n} (\sigma_3)_{(n)}\ \gamma(u),
\aeqno c\cr
\ &\cr
\gamma(-u)\ \ \ &=&\prod_{n} (\Omega)_{(n)}\ \gamma(u),\aeqno d\cr
\qqq
\vskip-\baselineskip\no
with $\Omega$ the generator of the Weyl group.
These properties come from \pref{blocks} specialized to
the gauge transformations that send
connections $A^{01}_u$ to $A^{01}_{u'}$.
Conversely, every holomorphic map $\gamma$ on
$\NC$ satisfying (\ref{cond},a-d), defines by \pref{blocks} and \pref{blocks1}
a unique functional $\Psi$ on the open dense
stratum $\CA_0^{01}$ but this {\bf need not} extend
to a holomorphic functional on the whole
of $\CA^{01}$. We explain now the necessary and sufficient
conditions for the existence of such an extension.

Define
$
g_u=\exp({\textstyle{-1\over2u}}\sigma_+).
$
Then
\qq\nonumber
{}^{h_\alpha} M_{u}d\bar z\equiv
{}^{h_\alpha g_u}A_u^{01}{-\hskip -2mm\longrightarrow}
\hskip -0.8cm{}_{_{\scriptstyle{u\to0}}}\hskip 0.3cm \hat A_\alpha^{01}.
\qqq
{}From now on we will use a realization of spaces
$V_{j_n}$ in terms of polynomials of degree at most
$2j_n$ on which elements of $SL(2,\NC)$ act by fractional
transformations. Using this realization we have (from \pref{blocks})
\qq
&&\bigotimes_n(\ee^{M_{u}
(\bar z_n- z_n)}h_\alpha^{-1}(z_n))_{(n)}
\ \Psi({}^{h_\alpha}M_{u}d\bar z)\big((v_n)\big)\cr\cr
&=&e^{\pi k(u+\alpha)^2/\tau_2}\hspace{0.13 cm}
\big(\prod_ne^{-(\alpha-\bar\alpha) z_nj_n/\tau_2}\big)
\hs{.13cm}(2u)^{-J}\hs{0.15 cm}
\cr&&
\gamma(u+\alpha)
\left((e^{(\alpha-\bar\alpha)z_n/\tau_2}+
2e^{(\alpha-\bar\alpha)z_n/\tau_2}uv_n)\right).
\label{alpha1}
\qqq
Now the left hand side is holomorphic in $u$ whereas the right hand
side, because of the presence of the $u^{-J}$ factor, is not holomorphic
unless
\qq
\partial_u^{l_0}\partial_{v_1}^{l_1}\cdots\partial_{v_N}^{l_N}
\gamma(u)({\underline v})\vert_
{u=\alpha\hfill\atop v_i=\exp[(\alpha-\bar\alpha)z_i/\tau_2]}=0
\label{reg1}
\qqq
\hspace*{1cm}for every $N+1$-tuple of non negative integers
$L\equiv(l_i)$ such that $\displaystyle|L|\equiv\sum_{i=0}^{N} l_i < J$.
\vskip 0.55 cm

Conversely, one may see that if \pref{reg1} holds for
$\alpha=0,1/2,\tau/2,(\tau+1)/2$ then
$\Psi$ can be extended to the four codimension $1$
strata $\CA^{01}_{(\alpha,0)}$.
Properties of the stratification allow now to
apply inductively Hartogs theorem and to extend $\Psi$, as a
holomorphic functional, to all other higher codimension strata.
In this way we obtain a quantum state of the theory.

\vskip 2mm
\no{\bf Theorem:} There is a one to one correspondence between
the space of states of Chern-Simons theory in the presence
of Wilson lines
and the space $W_N^{fr}(\tau,\un z)$ of holomorphic
functions $\gamma:\NC\rightarrow\bigotimes_nV_{j_n}$
satisfying (\ref{cond},a-d) and \pref{reg1} for
$\alpha=0,1/2,\tau/2,(\tau+1)/2$.
\vskip 2mm

As an illustration, we will analyze the two simplest cases of zero and
one insertions.
\vskip 1mm

\no\un{\it States with zero insertions.}

{}From (\ref{cond},a-d), the zero-points states are simply
the even theta-functions of degree $2k$, since for $J=0$ condition (\ref{reg1})
plays no role.
The dimension
of the space of even theta-functions is $k+1$. It is spanned by
Kac-Moody characters
$\chi_{k,j}(\tau,\ee^{2\pi iu})$ with fixed $\tau$ and $j=0,1/2,\dots,k/2$.
\vskip 2mm
\no\un{\it States with one insertion.}

In this case the states with one insertion of spin $j\in\NN$ are
\qq\nonumber
\gamma(u)(v)=\vartheta(u)v^j,
\qqq
where $\vartheta$ is a theta-function of degree $2k$ with
\qq
&&\vartheta(-u)=(-1)^j\vartheta(u)\alabel{1p}a\cr\cr
&&\partial_u^l\vartheta(u)|_{u=\alpha}=0 \hbox{\ for $l<j$ and
$\alpha=0$,$1\over 2$,$\tau\over 2$,$\tau+1\over 2$.}
\aeqno b\cr
\qqq
To study the dimension $d_j$ of the space of solutions of
(\ref{1p},a-b) we shall first focus on the case
of even spin, $j\in 2\NN$.
In that case (\ref{1p},a) requires even theta-functions for which:
\qq
&\ee^{-4 \pi i ku\textstyle{\alpha-\bar\alpha\over\tau-\bar\tau}}
\vartheta(u+\alpha)=
\ee^{4 \pi i ku\textstyle{\alpha-\bar\alpha\over\tau-\bar\tau}}
\vartheta(-u+\alpha)&\hbox{\hskip 1cm$\alpha=0$,
$1\over 2$,$\tau\over 2$,$\tau+1\over 2$.}\label{evthe}
\qqq
As a consequence of \pref{evthe}, vanishing of
$\partial_u^l\vartheta(u)|_{u=\alpha}$ for $l\leq 2n\in 2\NN$ implies
also vanishing of the same expression for $l=2n+1$. Thus we are left
with $2j$ linear conditions on the $k+1$-dimensional space of even
theta-functions and we obtain a lower bound on the dimension of
the space of solutions:
\qq
d_j\geq k-2j+1.\label{lbound}
\qqq

To obtain an upper bound for $d_j$ we shall use the fact
that the sum of multiplicities of zeros of the theta function
in any fundamental cell is $2k$. Now, if we have $d_j$ independent
even theta-functions with common zeros of total multiplicity $4j\leq 2k$,
by a linear combination of them we can obtain another
theta-function whose zeros have multiplicity at least $4j+2(d_j-1)$.
Since the latter has to be $\leq2k$, it follows that $d_j\geq k-2j+1$
and we obtain the expected result
\qq
d_j=\cases{
\matrix{k-2j+1 \hfill&&\hbox{for \ }j\leq k/2,\hfill\cr
0\hfill&&\hbox{otherwise}.\hfill}}\label{dim}
\qqq

The case of odd spin $j$ can be treated in a similar way. Now
we have $k-1$ independent odd theta-functions
and (\ref{1p},b) gives $2j-2$ independent conditions. Finally
we obtain the same expression (\ref{dim}) for the dimension.
\vskip 3mm
It can be shown (see \cite{fg}) that spaces
$W_N^{fr}(\tau,\un z)$ form a bundle
$W_N^{fr}$ over $\NH\times\NC^N$.
The modular group $\Gamma_N=SL(2,\NZ)\hs{0.08 cm}{\dl n}\NZ^{2N}$
acts on the base and its action can be lifted to $W_N^{fr}$.
Taking the quotient $W_N^{fr}/\Gamma_N$ we obtain a new
bundle over the moduli space of punctured torus.

Bundle $W_N^{fr}$ is endowed with a flat connection,
generalization of the Knizhnik-Zamolodchi\-kov connection \cite{KnizhZamo}
to the toroidal geometry \cite{Denis1}. The connection allows to compare
states at different points of the base. It is not preserved by the
modular group but gives rise to a projective
connection on $W_N^{fr}/\Gamma_N$.

By studying the behavior of spaces $W_N^{fr}$ on the boundary
of the moduli space of punctured tori, we were able to prove the
Verlinde formula for their dimensions modulo the result
that there are no toroidal states if one of the spins $j_n>k/2$.
The latter fact was proven in \cite{GK} for the spherical states
and in the toroidal case should follow by studying the spherical
limit $\tau\rightarrow\infty$ which requires, however,
some additional work, the goal of a future
research.

\end{document}